\documentclass{wrup}

\def\stot{\mbox{$\sigma_{\rm tot}^{\gamma p}$}}

\begin{document}

\title{\boldmath $F_2^{\rm p}$ at low $Q^2$
       and the total $\gamma p$ cross section at HERA}

\author{S. Levonian}

\address{DESY, Hamburg, Germany\\ 
E-mail: levonian@mail.desy.de}  

\twocolumn[\maketitle\abstract{
New precise measurements of the neutral current cross section at HERA 
are presented in different $Q^2$ regimes.
In the photoproduction limit the total $\gamma p$ cross section is found
to be in agreement with the universal soft Pomeron prediction.
At medium scales $1.5 \leq Q^2 \leq 150$ GeV$^2$ both the $F_2$ and the $F_L$
structure functions are determined. These data are well described 
by NLO pQCD. The gluon density in the proton has been extracted from the
scaling violation of $F_2$ at low $x$.
The transition between the two regimes occurs at around $Q^2 \simeq 1$
GeV$^2$.}] 
%and its underlying dynamics is still a challenge for theory.}]

\section{Introduction}
%=======================

Half a century of the extensive studies of strong interactions
resulted in two different approaches to the problem:
Reggeon field theory (RFT)\cite{RFT} and QCD.
RFT, the $S$-matrix theory based on the most general physical principles,
has proven to be very successful in describing soft peripheral processes
($t/s \ll 1$) at high energies ($s \to \infty$). There is however no
microscopic picture of underlying dynamics in it.
On the other hand QCD, being {\em the} theory of strong interactions,
has technical problems in the non-perturbative regime. pQCD is applicable
to hard processes only.
Lepton-proton scattering experiments at HERA have the unique possibility 
to contribute to the successful merging of those two approaches
by performing the scan over the large available range of the photon virtuality, 
$Q^2$, and studying the interplay of short and long distance physics.

In this talk  the following questions are discussed, 
using high statistics Neutral Current (NC) data 
recorded in the years 1994 to 1997 with the H1 and ZEUS apparatus
in $e^+p$ collisions at $\sqrt{s}=300$ GeV: 
\begin{enumerate}
  \item How far up in $Q^2$ can one get with Regge theory
        starting from photoproduction?
  \item How far down in $Q^2$ can one go with pQCD?
  \item Where in $Q^2$ is the transition region? 
\end{enumerate}

\section{Photoproduction limit}
%===============================

The total $\gamma p$ cross section is measured at HERA by detecting
scattered positrons under very small angles $\theta\!<\!5$ mrad with respect to
the incoming $e^+$ beam,
in a special calorimeter installed in the tunnel.
This ensures $Q^2 \!< 0.02$ GeV$^2$ (with the average value of $10^{-4}$ GeV$^2$)
and justifies the use of pure transverse photon flux in the equivalent photon 
approximation relating the $ep$ cross section with $\stot$.
Major systematics is approximately equally shared between the positron detector
acceptance uncertainty and the precision of the hadronic final state modelling.

\begin{figure}[hb]
\epsfxsize195pt
\figurebox{}{}{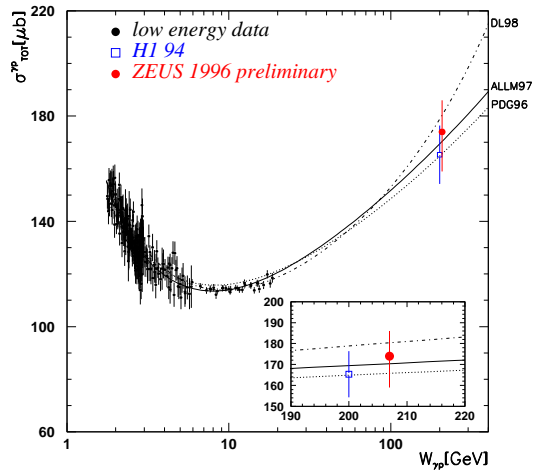}
\caption{The total photon-proton cross section as a function of
         centre-of-mass energy. The insert magnifies HERA results.}
\label{fig:stot}
\end{figure}

New measurement of the total $\gamma p$ cross section  
at average centre-of-mass energy $W_{\gamma p}=207$ GeV
has been presented by the ZEUS collaboration~\cite{Zstot}. 
Their result of $\stot = 172 \pm 1(stat.)^{+13}_{-15}(syst.)~\mu$b 
is shown in Fig.~\ref{fig:stot} together with the 
published H1 measurement~\cite{Hstot} and the low-energy data.
Also shown are three Regge motivated parameterizations~\cite{Regge}. In two of them
the high energy behavior of $\stot$ is driven by the universal soft Pomeron, while
the DL98 parameterization has additionally a hard Pomeron term.

\section{NC cross section in DIS regime}
%========================================

\begin{figure}[hb]
%\epsfxsize170pt
\epsfig{file=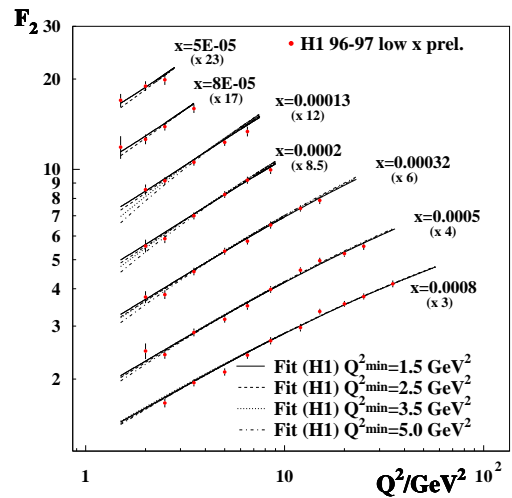,width=183pt}
\epsfig{file=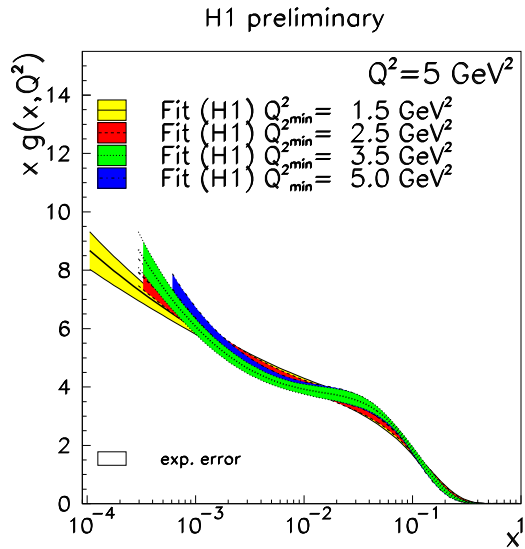,width=183pt}
\caption{Effect of the $Q^2_{min}$ cut applied in DGLAP QCD fit to the H1 data on 
         a) the structure function $F_2$ at low $Q^2$ and
         b) the gluon distribution at $Q^2=5$ GeV$^2$.}
\label{fig:gluon}

\end{figure}

New measurements of the deep inelastic NC cross section are available from both
the H1~\cite{HF2} and ZEUS~\cite{ZF2med} collaborations. 
High statistics data span the kinematic region of $1.5 \leq Q^2 \leq 150$ GeV$^2$ 
and Bjorken-$x$  values $0.00003 \leq x \leq 0.2$.
Using improved detection capabilities and increased HERA luminosity a high
precision of typically 3$\%$ is achieved.
This allowed, for the first time, to perform NLO QCD analysis of the 
inclusive cross section measurements using H1 data alone~\cite{HF2}.
The gluon density at low $x$ has been determined from the large positive
scaling violation of $F_2$ (see Fig.~\ref{fig:gluon}).
It was found, that NLO QCD fit describes all low $x$ data well.
Some dependence is observed however of the fit parameters on the value of
$Q^2_{min}$ -- the minimum $Q^2$ of the H1 data used in the fit. This is directly
reflected in the steepness of the gluon density, as seen in Fig.~\ref{fig:gluon},b.
  
\begin{figure}[tb]
\epsfxsize200pt
\figurebox{}{}{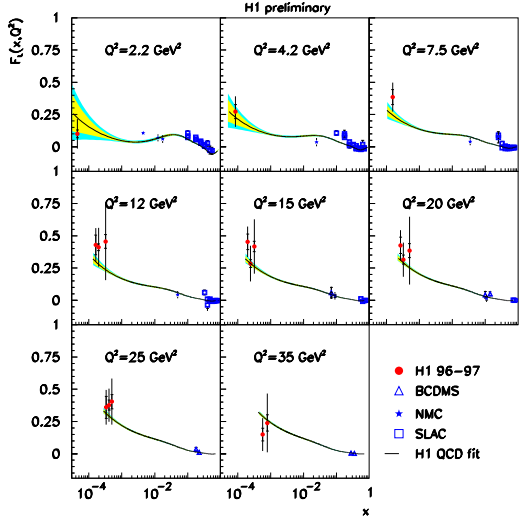}
\caption{The longitudinal structure function $F_L(x,Q^2)$.
         The error bands are due to the experimental (inner) and model (outer)
         uncertainty of the $F_L$ calculation using NLO QCD fit to the H1 data
         for $y<0.35$ and $Q^2>3.5$ GeV$^2$.}
\label{fig:FL}
\end{figure}

In order to reduce such `flexibility' of the QCD fit additional constraints
may be imposed.
Potentially powerful is the longitudinal structure function $F_L$,
which contains independent information about gluon distribution.
$F_L(x,Q^2)$ as determined using two different methods~\cite{HF2} is shown in 
Fig.~\ref{fig:FL} together with the fixed target results. The increase of
$F_L(x,Q^2)$ towards low $x$ reflects the rise of the gluon momentum distribution
and is consistently described by NLO QCD fit.

\section{\mbox{Transition region: closing the gap}}
%==================================================

To summarize, HERA has verified that RFT works in photoproduction ($\stot$).
New precision results also demonstrate that in DIS regime pQCD describes
inclusive NC cross section ($F_2, F_L$). But where is the transition between
the two? And how does it happen?


\begin{figure}[tb]
\epsfxsize190pt
\figurebox{}{}{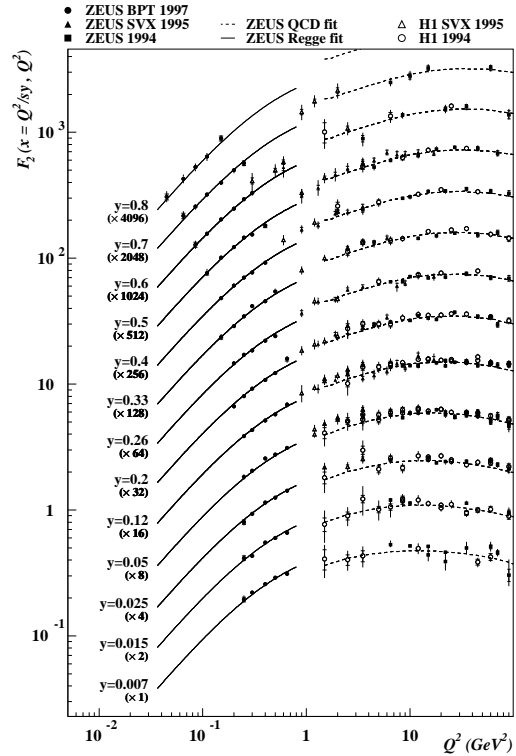}
\caption{Measured $F_2(Q^2)$ in bins of $y$.}
\label{fig:BPT}
\end{figure}

\noindent
New low $Q^2$ ZEUS data~\cite{ZF2} at $0.045 \leq Q^2 \leq 0.65$ GeV$^2$
almost completely closed the gap between photoproduction and DIS.
They are shown in Fig.~\ref{fig:BPT} together with
previous HERA measurements at higher $Q^2$.
It is seen that low $Q^2$ points are described adequately by the 
Regge motivated fit (solid lines) while pQCD fits the data above 1.5 GeV$^2$
(dashed lines).
The data exhibit a smooth transition at around $Q^2 \approx 1$ GeV$^2$
while the matching between the two theoretical fits is not perfect yet.

\section{Conclusions}
%====================

New precise HERA data have been used to study how the properties
of strong interactions evolves with $Q^2$.
In the photoproduction limit $\stot$ ~exhibits mild rise similar to that of
hadron-hadron scattering. 
It is well described by the conventional Regge theory with universal soft Pomeron.
Regge parameterization also describes the data in the low $Q^2<0.7$ GeV$^2$ region. 
In DIS regime NLO QCD is able to describe $F_2$ data all the way down to 
%surprisingly low scale, 
$Q^2 \simeq 1.5 $ GeV$^2$.
A smooth transition from partonic to hadronic
degrees of freedom occurs at around $Q^2 \simeq 1$ GeV$^2$.
The details of the underlying dynamics is still 
a challenge for theory.

\end{document}